# Spatial-domain interactions between ultra-weak optical beams

Utsab Khadka,[1,*] Jiteng Sheng,[1] and Min Xiao[1,2,§]

[1]*Department of Physics, University of Arkansas, Fayetteville, Arkansas 72701, USA*

[2]*National Laboratory of Solid State Microstructures and Department of Physics, Nanjing University, Nanjing 210093, China*

*ukhadka@uark.edu

§mxiao@uark.edu

**Abstract:** We have observed the spatial interactions between two ultra-weak optical beams that are initially collinear and non-overlapping. The weak beams are steered towards each other by a spatially varying cross-Kerr refractive index waveguide written by a strong laser beam in a three-level atomic medium utilizing quantum coherence. After being brought together, the weak beams show controllable phase-dependent outcomes. This is the first observation of soliton-like interactions between weak beams and can be useful for all-optically tunable beam-combining, switching and gates for weak photonic signals.

It is well known that two optical fields that are spatially apart can interact with each other in a nonlinear optical medium, and that the interaction can be tuned via the relative phase between the fields[1-9]. This effect has been demonstrated commonly in soliton collision experiments where, depending on the relative phase between the solitons, different outcomes are achieved such as fusion and repulsion. These experimental observations were made in photorefractive crystals as well as atomic vapors, which have different (quadratic and cubic) nonlinearities, respectively. In these experimental demonstrations, the solitons are achieved when the laser beams self-waveguide themselves. The underlying mechanism of self-waveguiding is self-focusing, which is a nonlinear effect arising due to an intensity-dependent refractive index n(I); because of the beam's Gaussian intensity distribution I(**r**), n(I) causes a lensing effect for the beam as it propagates through the nonlinear medium, thus overcoming the natural tendency of the beam to diverge[10-12]. The conditions for stability of solitons of different dimensions have been extensively investigated[13-14]. The self-induced nonlinearities require the beams to have large intensities, and often high-powered pulses with very narrow temporal and spatial widths are used. In a two-soliton interaction, the interference between the two fields causes the intensity in the region between them to vary with their relative phase difference. For instance in the in-phase case, constructive interference enhances the intensity and the nonlinear refractive index in this region, thus steering both solitons to this region and resulting in their fusion. Such phase-dependent outcomes for two beams that have an initial spatial separation can be important in constructing all-optical gates and switches for optical signals, as shown in the schematic in Fig. 1. Here, we ask the question: can similar spatial-domain interactions be induced between two beams having very weak intensities and insignificant self-induced nonlinearities? Such interactions will be important ingredients in communication and computation protocols for weak photonic signals. Recently, theoretical studies have predicted the generation of stable ultra-weak intensity solitons and their collisions in three- and four- level quantized systems[15-22], where the all-optical wave-guiding is achieved via quantum coherence effects induced by additional strong coupling beams.

In this Letter, we experimentally demonstrate the phase-dependent interactions between two initially spatially separate optical fields having very weak intensities. Here, the underlying mechanism is a spatially varying cross-Kerr refractive index profile for the weak beams induced by a strong coupling beam that is initially partially overlapped with both of the weak signal beams inside a vapor cell containing three-level rubidium atoms and utilizing quantum coherence to generate a common, tunable all-optical waveguide. In this way, by utilizing the cross-Kerr effect, we relax the requirement for the two signal fields themselves to have large intensities in order to achieve the spatial refractive index gradient. First, we will describe the experimental setup and the mechanism of the coupling-beam-induced waveguide. Numerical results will be presented to show that a small set of parameters and initial conditions exist that yield the required spatial refractive index and stable wave-guiding of the weak probe beams in the Doppler-broadened atomic medium. We will then demonstrate that once both the weak signal beams are steered into the common all-optical waveguide, the resulting interaction between them can be controlled via the relative phase-difference between them. The interaction shown in Fig. 1 is demonstrated between these two weak beams; in addition, we also show that in this system, there are more tunable parameters compared to the scheme that uses two high-powered self-focusing fields.

The experimental setup and the atomic system are shown in Fig. 2. Two weak probe beams $E_1$ and $E_1'$ are derived from the same diode laser. The output of the diode laser is first fed into a single-mode polarization maintaining fiber (not shown in figure) for mode-cleaning. The strong coupling beam $E_c$ is from a Ti:Sa ring laser. All three beams are continuous-wave (cw), nearly collimated, and have Gaussian spatial profiles.

Beams $E_1$ and $E_1'$ are nearly collinear with a vertical separation between their centroids and negligible overlap. The beam $E_c$ counterpropagates with $E_1$ and $E_1'$, such that $E_c$'s centroid is in the middle of the centroids of $E_1$ and $E_1'$, and $E_c$ has overlap with both $E_1$ and $E_1'$. That is, beams $E_1$ and $E_1'$ lie on the opposite sides of counterpropagating beam $E_c$, and the centroids of the three beams lie in the x-z plane. $E_1$ and $E_1'$ are linearly polarized in the plane containing them, while $E_c$ has a linear polarization orthogonal to $E_1$ and $E_1'$. The strengths of $E_1$ and $E_1'$ can be controlled independently by the various half-wave plates and polarization beam-splitters, and together by the neutral density attenuator wheel. One of the mirrors M2 in the path of $E_1'$ is piezo-actuated, allowing control of the relative phase $\Delta\Phi_1$ between $E_1$ and $E_1'$. Another mirror on $E_1'$ s path (not shown in figure) is placed on a micro-meter translational stage, such that the relative separation between the fields can be easily tuned.

Various lenses (not shown in the figure) are used to control the widths and collimations of the fields. The average width of the coupling beam inside the vapor cell is $w_c$= 156 µm. The widths of the probe beams at the entrance of the vapor cell are $w_1$ = 133 µm and $w_1'$=148 µm, and the distance between their centers is 120 µm. When propagating in free space, these beams slowly diverge and by the time they travel 325 mm to the charge coupled device camera CCD2, where they are imaged without using a lens, their widths are $w_1$ = 670 µm and $w_1'$ = 600 µm. In order to prevent the overlap of these probe beams at CCD2, they are aligned with a small angle between them so that at CCD2, the separation between their centers is larger than their widths (Fig. 4(a)).

The three fields pass through a Rb vapor cell that is heated to 95 °C by a heating coil. The vapor cell is 7.5 cm long, and 3.5 cm of the cell's central portion is accessible for fluorescence imaging. The fields $E_1$ and $E_1'$ are nearly resonant with the $D_2$ transition (~780.23 nm). They have a frequency detuning of $\Delta\nu_1$ = 400 MHz towards the blue side of the F=3 → F' = 4 transition of $^{85}$Rb isotope. The coupling beam $E_c$ drives the $5P_{3/2} \to 5D_{5/2}$ transition (~775.98 nm), and its frequency detuning $\Delta\nu_2$ can be tuned in 10 MHz increments. The transverse spatial profiles (x-y dimension) of the transmitted probe beams are imaged at CCD2, which is 25 cm away from the Rb cell's exit. The counter-propagating beam geometry effectively isolates the strong coupling beam from the weak probe beams before they reach CCD2, allowing images of the weak fields to be taken at the CCD2 with highest sensitivity setting while maintaining minimal background noises. The transmitted field $E_1$ is also monitored by a photodiode for spectral characterization.

When the camera is placed at position CCD1, it takes images of the x-z dimension of the beams inside the vapor cell via fluorescence from the side of the Rb cell. The fluorescence is imaged onto CCD1 by two cylindrical lenses CYLx and CYLz with focal lengths 10 cm each, which are positioned such that the x- and z- dimensions are magnified by factors of 4 and ¼ at CCD1, respectively. This way, a significant axial length of the beams within the vapor cell (almost 25 mm) can be imaged in a single image while still maintaining a good resolution of the transverse (x) dimension (about 1 mm), at the 7.04 mm x 5.28 mm CCD surface. This circumvents the need to take several axial images and patching them together per image for high resolution of the transverse dimension. Spherical lenses cannot provide this feature, since they magnify or demagnify both dimensions equally. CYLx, CYLz and CCD1 are each placed in three-dimensional micrometer-precision stages with translational and rotational degrees of freedom to facilitate the imaging process. Furthermore, the beams $E_1$ and $E_1'$ are linearly polarized in the x-z plane so as to maximize the dipole-scattered radiation pattern at CCD1.

In the three-level ladder-type atomic medium, a strong coupling beam alters the absorption and dispersion of a weak probe beam via quantum coherence. When both one-photon and two-photon resonances (TPR) are met, the atomic medium is rendered transparent for the probe beam by virtue of electromagnetically induced transparency (EIT). Within the spectral window of EIT, the transmission of the probe beam increases, and it also experiences a rapidly varying refractive index. The value of the refractive index can be controlled by the coupling beam's intensity $I_c$ and the two-photon detuning. The modified absorption and refractive index of the probe beam in this three-level Doppler-broadened atomic medium are given by the imaginary and real parts, respectively, of the complex susceptibility $\chi$ [23, 24].

The dependence of the susceptibility on $\Omega_c^2$, where $\Omega_c$ is the coupling beam's Rabi frequency, and the Gaussian spatial distribution of $I_c(\mathbf{r})$, means that the weak probe beam experiences a spatially varying refractive index $n(\mathbf{r})$ which can give rise to lensing and waveguiding behaviors. Such cross-Kerr induced focusing and defocusing for perfectly overlapped probe and coupling beams in the ladder-type configuration with counterpropagating geometry was reported in Ref. 25. The variation of absorption and refractive index as a function of the coupling beam's intensity in an EIT medium have been well characterized before, including in the ladder-type configuration[23-25].

In our setup the coupling beam $\mathbf{E_c}$, which is on the order of $10^5$ more intense than the probe beams, has a rapidly varying spatial profile. The probe beams are placed at the opposite sides of this Gaussian intensity distribution. In order to find the correct profile for n($\mathbf{r}$) that yields stable waveguiding of the weak probe beams along the axis of the coupling beam, the correct combination of parameters ($I_c$, $\Delta v_1$, $\Delta v_2$, cell temperature,…) has to be found. We explored the large parameter space both experimentally and numerically. We find that the correct shape of n($\mathbf{r}$) supporting waveguiding, as well as the stability of the waveguided beams, are very sensitive to the parameters and initial geometrical conditions.

Numerically calculated transverse refractive index profiles for various parameters are shown in Fig. 3a. To get these results, the density matrix for the three-level system is solved, and then numerically integrated over all the atomic velocity classes to account for Doppler broadening. It is clear that only certain parameters will yield index profiles that support confinement of the weak beams. The parameters giving rise to the refractive index profile in Fig. 3 (a(iii)), which are similar to the parameters used in the experiment, are used in tracing the trajectories of the probe beam $\mathbf{E_1'}$ for different initial conditions (Fig. 3(b)). The trajectories are found by extremizing the time taken by the rays to traverse through the medium, i.e. by using Fermat's principle. The probe beam is waveguided for only a small range of initial transverse position $x_0$ and angle $\theta_0$ with respect to the coupling beam's axis. Fig. 3(b) shows numerical solutions demonstrating some values of $x_0$ leading to waveguiding, as well as some that do not support the waveguiding. Here, various trajectories of one probe beam are traced; the other beam's trajectories show similar dependence on the parameters and initial conditions due to the radial symmetry of the induced refractive index. We have also numerically seen that the waveguiding occurs for only a small range of initial launch angle $\theta_0$.

Once the two weak fields (~400 nW each) are steered inside this common all-optical waveguide, the interaction between them becomes dependent on their relative phase difference $\Delta\Phi_1$. When $\Delta\Phi_1 = 0$, the intensity in the central region of the waveguide becomes maximum (Fig. 4(b)). When $\Delta\Phi_1 = \pi$, the two fields interfere destructively and the central region of the waveguide remains dark (Fig. 4(c)). This is equivalent to the interaction of Fig. 1, which was previously experimentally demonstrated between two strong self-guided beams, and now demonstrated between two ultra-weak beams using quantum coherence. Furthermore, the output state in the current case has more tunability. For the in-phase case (Fig. 4(b)), the output intensity of the central bright component can be all-optically tuned via the intensity of the coupling beam (Fig. 4(d)). This is because in this system we not only modify the refractive index, but the transparency of the medium itself. In the previous demonstrations of two-beam fusion using self-induced nonlinearities, the output intensity cannot be tuned since the fusion is critically dependent not only on the relative phase, but also on the signal beam intensities themselves which cannot be reduced otherwise the self-induced nonlinearity will disappear.

Another novel feature of this system is that for the out-of-phase case, the two signal beams do not spatially deflect away; instead, due to the attractive central potential induced by the coupling beam, both beams are guided tightly to the central axis, while the axis itself remains dark due to destructive interference. This opens the room for generating dark vortices with enhanced depth. This feature is possible because in this system, the strength of the attraction is controlled externally by the frequency detuning and intensity of the coupling beam, and not by the intensities of the signal fields themselves.

For large two-photon detunings ($\Delta\nu_2 = \pm 250$ MHz) the effects of EIT disappear and the transmission of the resonant probe beams through the vapor cell decreases sharply. In this case, we imaged the incoherent fluorescence signal through one side of the vapor cell. In order to increase fluorescence signal for imaging, the probe beam's powers were increased to 400 μW each. Even in these large-detuned cases, the spatially varying refractive index due to $I_c$ presents itself as an attractive or a repulsive potential acting on the weak probe fields. The paths of the two resonant probe beams in the absence of the coupling beam are shown in the image taken by CCD1 in Fig. 5 (a). The beams look overlapped because in the region between, the intensities due to the fluorescence caused by each beam add up. Note the different scales of the x- and z-dimensions. For a positive (negative) $\Delta\nu_1$, we observe that a positive (negative) $\Delta\nu_2$ pulls both probe beams towards the wave-guide center (Fig. 5b) while a negative (positive) $\Delta v_2$ pushes the weak beams further apart (Fig. 5c). Transverse cross-sections of each image at a fixed longitudinal position are shown in Fig. 5d. While the enhancement and decrease of the resultant intensity in the central axis is apparent from these images and traces, the contrast is degraded due to the large frequency detuning and thus weaker atomic coherence, and also due to background scattering by the windows of the vapor cell. On the other hand, this noise would have been overwhelming had we used the lambda-type atomic configuration, since in this case the strong coupling beam also has access to the ground state atoms and causes single resonance fluorescence. One of the main motivations for using the ladder-type scheme is that the strong coupling beam drives the transition between two excited states which have no atomic population in equilibrium, and thus does not contribute to fluorescence unless the probe beam is present. As a result, the measurable image signal-to-noise ratio is much higher in this three-level ladder-type atomic configuration.

We have thus utilized quantum coherence induced by a strong coupling beam in a three-level atomic system to observe waveguiding and spatial interactions between two ultra-weak beams, for resonance as well as off-resonance of the two-photon frequency detuning. In the on-resonance case, where absorption is suppressed due to EIT and we measured the transverse profile of the transmitted probe beams, we observed phase-dependent interactions akin to soliton combination and repulsion. In the off-resonance case, where incoherent scattering is large, we observed longitudinal side-ways fluorescence images. We have shown that the system has a large set of tunable parameters (single-photon and two-photon frequency detunings, coupling beam's power, relative phase between the probe beams), and that it allows all-optically tunable waveguiding, combining, switching and routing of ultra-weak beams, as well as controlling interaction between them.

While the current system shares a common capability with a traditional beam splitter, i.e. combination of two fields and a phase-dependent output, the current system has some novel features. The first is the combination geometry- this system combines two collinear and non-overlapping beams by using all-optical cross-Kerr nonlinearity. Second, while the mechanism of beam combination in a traditional beam-splitter is fixed transmission and reflection, the mechanism in the current system is tunable transmission and refraction. Third, when the two input beams are in phase, the output intensity in the current system can be varied by tuning the coupling beam's intensity since along with the real part of the refractive index, the imaginary part is also tunable. Due to these novel features, the system can also be thought of as an all-optically reconfigurable beam-combiner for collinear fields, and the new geometry and transistor-like variable output state can be useful in integrated photonic circuit geometries. Furthermore, since EIT is a natural test bed for slow light as well as stored light[22-33], it will be useful to extend this system to study quantum memory and quantum logic gates involving two ultra-weak fields having phase-dependent transverse spatial interactions.

**Acknowledgements**

U.K. thanks G. Salamo and S. Singh for their invaluable discussions on experimental imaging and numerical ray-tracing methods. U.K. was supported by the Raymond H. Hughes Research Fellowship.

**Figure Captions:**

**Fig. 1.** Schematics of phase-dependent spatial interactions between two optical fields inside a nonlinear optical medium (green shaded region). The two fields propagate along **z**, and have an initial separation along **x**. The output state depends on the relative phase $\Delta\Phi$ between the two fields.

**Fig. 2.** Atomic system and simplified experimental setup. Att = variable neutral-density attenuator wheel, (P)BS = (polarizing) beam splitting cube, H = half-wave plate, CYL = cylindrical lens, CCD = charge-coupled device camera, PD = fast photodiode, M = mirror. The inset shows the initial relative orientations between the three beams.

**Fig. 3.** Numerical calculations: (a) Transverse refractive index for $\Delta v_1 = +300$ MHz, $\Omega_c = 250$ MHz, $w_c = 156$ μm, (i) $\Delta v_2 = 0$ MHz (ii) $\Delta v_2 = +15$ MHz (iii) $\Delta v_2 = +25$ MHz (iv) $\Delta v_2 = +50$ MHz (v) $\Delta v_2 = +250$ MHz. (b) Probe beam trajectories for different initial transverse positions $x_0 = [5, 20, 35, 50, 60, 100, 200, 250, 270, 280, 285, 300]$ μm and transverse refractive index profile (a)(iii).

**Fig. 4.** Two-dimensional transverse images taken by CCD2 when two-photon detuning is nearly resonant. Cell temperature = 95 °C, $\Delta v_1 = +300$ MHz, $\Delta v_2 = +10$ MHz. Beam powers are measured before Rb cell. $P_1 = P_1' = 400$ nW. (a) $P_2 = 0$ mW, (b) $P_2 = 115$ mW, $\Delta\Phi_1 = 0$, (c) $P_2 = 115$ mW, $\Delta\Phi_1 = \pi$. In (d), the in-phase condition similar to Fig. 3 (b) is used, and the peak intensity of the central fused component is measured as a function of the coupling beam's power $P_2$.

**Fig. 5.** Longitudinal images taken by CCD1 when two-photon detuning is off-resonant. Cell temperature = 95 °C, $\Delta v_1 = +400$ MHz. Beam powers are measured before Rb cell. $P1 = P1' = 400$ μW. Note the different scaling of the axes: **x** is in μm but **z** is in cm, achieved by the specially designed imaging. (a) $P_2 = 0$ mW, (b) $P_2 = 90$ mW, $\Delta v_2 = +250$ MHz, attractive (c) $P_2 = 90$ mW, $\Delta v_2 = -250$ MHz, repulsive. In (d), (i), (ii) and (iii) are the one-dimensional cross-sections taken along **x** (at **z** = 2.35 cm) from the 2-D images shown in (a), (b) and (c) respectively.

**Figures:**

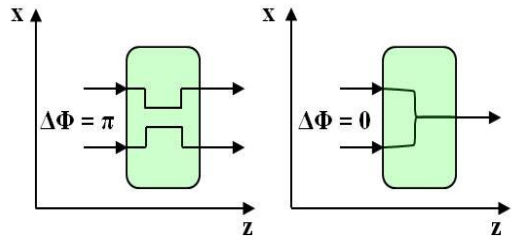

**Fig. 1. Khadka et al**

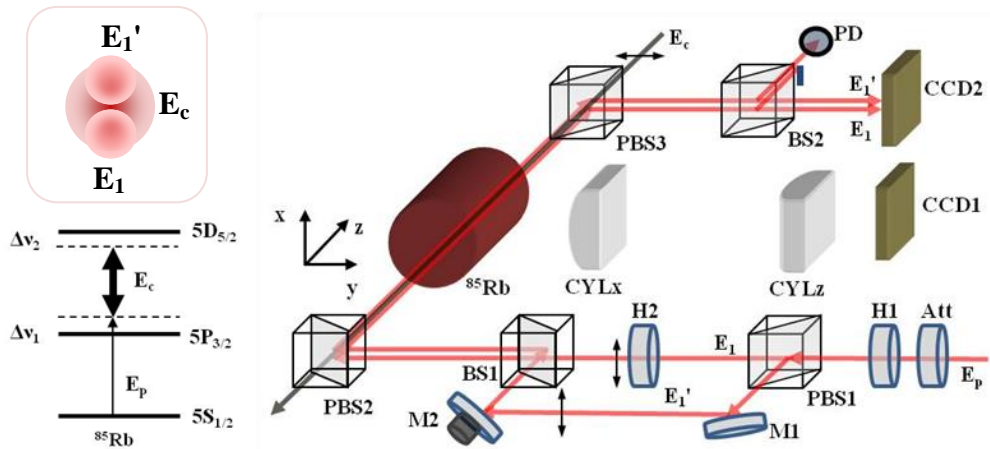

**Fig. 2. Khadka et al**

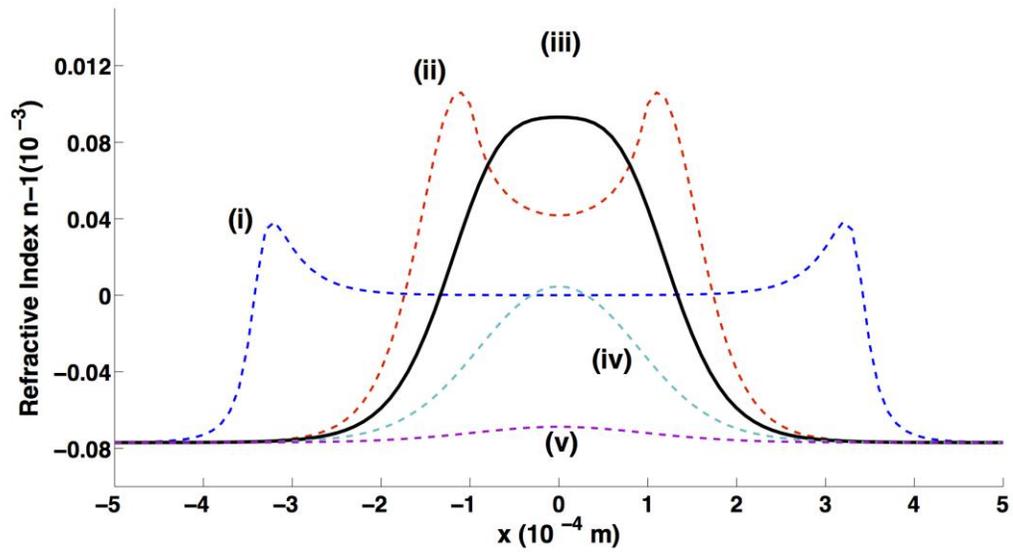

(a)

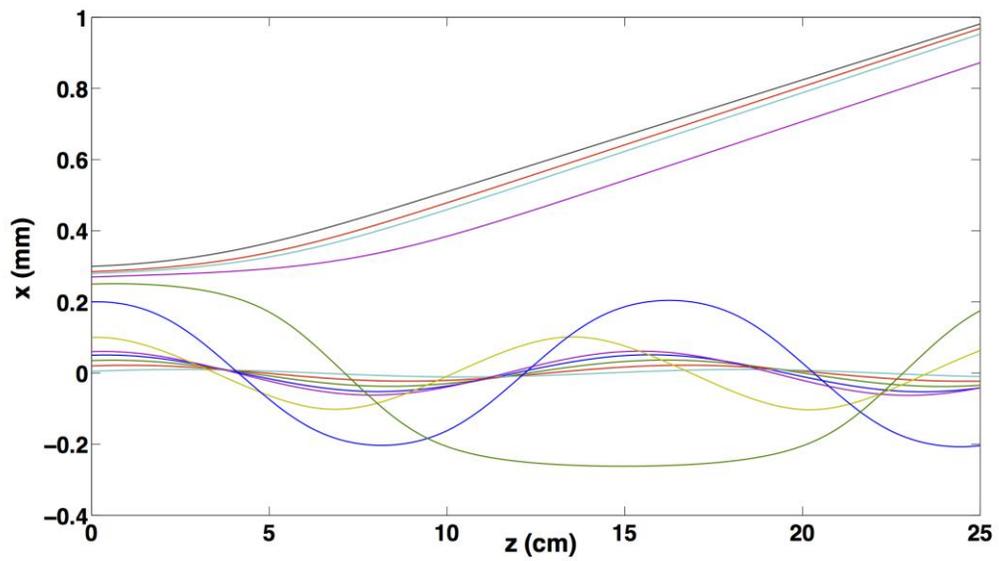

(b)

**Fig. 3. Khadka et al**

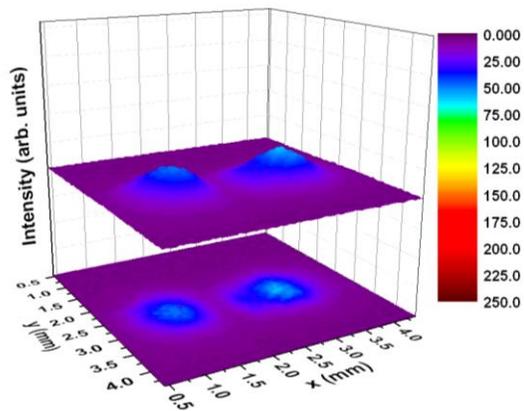

**(a)**

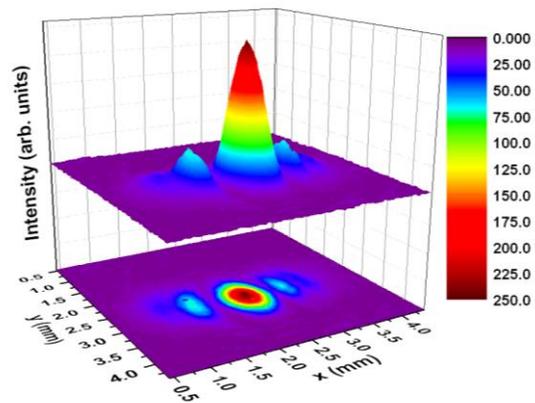

**(b)**

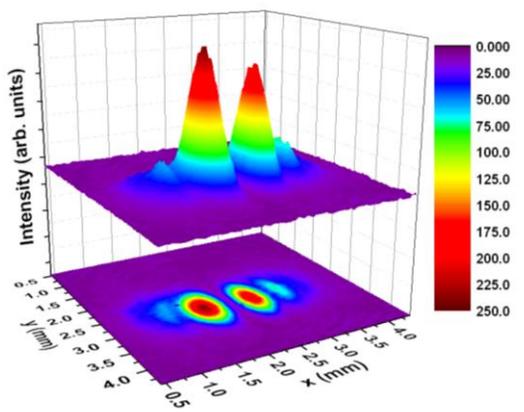

**(c)**

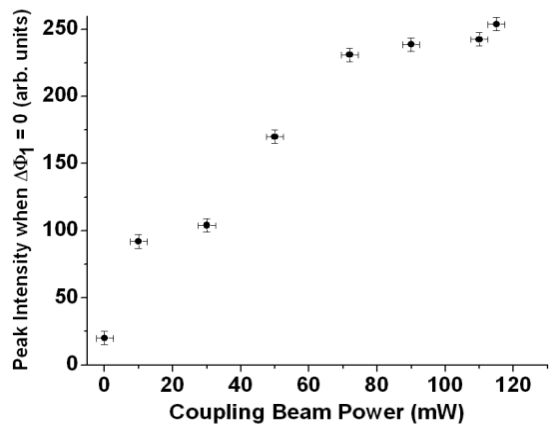

**(d)**

**Fig. 4. Khadka et al**

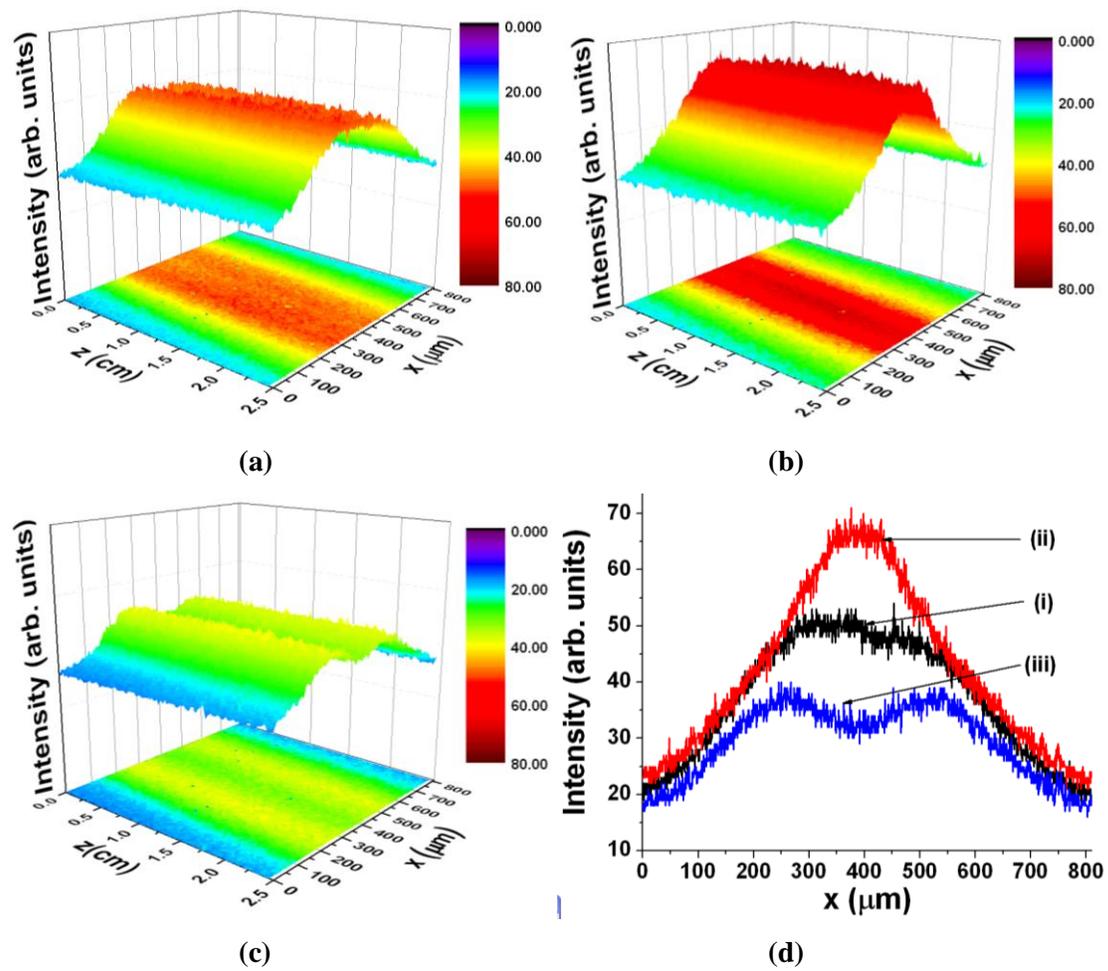

**Fig. 5. Khadka et al**